\documentclass[aps,prl,epsf,twocolumn,showpacs]{revtex4}
\usepackage{amsmath}
\usepackage{amssymb}
\usepackage{epsfig}
\usepackage{graphicx}
\usepackage{color}
\usepackage[normalem]{ulem}
\usepackage{tikz}
\usetikzlibrary{positioning}

\graphicspath {{./figures/}}
\makeatletter
\def\input@path{{./figures/}}
\makeatother

\begin{document}

\title{Universality from disorder in the random--bond Blume--Capel model}

\author{N.~G. Fytas$^1$}
\author{J. Zierenberg$^{2,3,4}$}
\author{P.~E. Theodorakis$^5$}
\author{M. Weigel$^1$}
\author{W. Janke$^2$}
\author{A. Malakis$^{1,6}$}

\affiliation{$^1$Applied Mathematics Research Centre, Coventry
University, Coventry, CV1 5FB, United Kingdom}
\affiliation{$^2$Institut f\"{u}r Theoretische Physik,
Universit\"{a}t Leipzig, Postfach 100 920, 04009 Leipzig, Germany}
\affiliation{$^3$Max Planck Institute for Dynamics and
Self-Organization, 37077 G\"{o}ttingen, Germany}
\affiliation{$^4$Bernstein Center for Computational Neuroscience,
37077 G\"{o}ttingen, Germany} \affiliation{$^5$Institute of
Physics, Polish Academy of Sciences, Al.\ Lotnik\'ow 32/46,
02-668, Warsaw, Poland} \affiliation{$^6$Department of Physics,
Section of Solid State Physics, University of Athens,
Panepistimiopolis, GR 15784 Zografou, Greece}

\date{\today}

\begin{abstract}
Using high-precision Monte Carlo simulations and finite-size scaling we study
the effect of quenched disorder in the exchange couplings on the Blume-Capel
model on the square lattice. The first-order transition for large crystal-field
coupling is softened to become continuous, with a divergent correlation length.
An analysis of the scaling of the correlation length as well as the
susceptibility and specific heat reveals that it belongs to the universality
class of the Ising model with additional logarithmic corrections observed for
the Ising model itself if coupled to weak disorder. While the leading scaling
behavior in the disordered system is therefore identical between the
second-order and first-order segments of the phase diagram of the pure model,
the finite-size scaling in the ex-first-order regime is affected by strong
transient effects with a crossover length scale $L^{\ast} \approx 32$ for the
chosen parameters.
\end{abstract}

 \maketitle

The effect of random disorder on phase transitions is one of the basic problems in
condensed-matter physics~\cite{young:book}. Examples include quantum Ising magnets
such as $\mathrm{LiHo_xY_{1-x}F_x}$ \cite{tabei2006,silevitch2007}, nematic liquid
crystals in porous media \cite{maritan1994}, noise in high-temperature
superconductors~\cite{carlson2006} and the anomalous Hall
effect~\cite{nagaosa2010}. Understanding random disorder in classical, equilibrium
systems is a crucial step towards solving the more involved problems in quantum
systems \cite{vojta2014}, for example many-body localization with programmable random
disorder \cite{smith2016}, and in non-equilibrium phase
transitions~\cite{barghathi2012}.

The case of weak disorder coupled to the energy density of systems with continuous
transitions is rather well understood: Uncorrelated disorder is relevant and leads to
new critical exponents if the specific-heat exponent $\alpha$ of the pure system is
positive, a rule known as the Harris criterion~\cite{harris74}. If long-range
correlations in the disorder are present, this rule can be generalized leading to
interesting
ramifications~\cite{weinrib:83,luck:93a,wj:04a,barghathi:14,fricke:17,fricke:17b}.
These effects, and in particular the marginal case of a vanishing specific-heat
exponent as present in the two-dimensional Ising model, are intriguing and have
attracted a large research effort over the past decades~\cite{dotsenko:83,
  shalaev:84, shankar:87,ludwig:88,wang:90,hasenbusch:08a,kenna:08,dotsenko:17}.

The situation is less clear for systems undergoing first-order phase transitions that
are much more common in nature The observation that formally $\nu = 1 / D$ and
$\alpha = 1$ for such systems in $D$ dimensions suggests that disorder is always
relevant in this case, and the general observation is that it indeed softens
transitions to become continuous~\cite{cardy:99a}. Such a rounding of discontinuities
has been rigorously established for systems in two dimensions~\cite{aizenman:89a},
but is believed to be more general -- a view that is supported by a mapping of the
problem onto the random-field model~\cite{hui:89a,berker93,cardy:97a}. This general
picture is commonly accepted, and similar phenomena are recently studied in quantum
systems~\cite{goswami:08,greenblatt:09,hrahsheh:12} and for non-equilibrium phase
transitions~\cite{vojta:06,martin:14}.  Still, a number of important questions have
not been answered in full generality: Is a finite strength of disorder required to
soften a first-order transition? Is there a divergent correlation length? What is the
universality class of the resulting continuous
transition~\cite{cardy:97a,bellafard:12,zhu:15}?

While a softening must occur for arbitrarily small disorder strength in two
dimensions~\cite{aizenman:89a,hui:89a,berker93}, the situation is less clear in
three dimensions~\cite{chatelain:01a,chatelain:05}, but in both cases one finds
divergent correlation lengths. The question of the universality class of
softened transitions is perhaps the most intriguing one. This has been studied
in some detail for the random-bond $q$-state Potts
model~\cite{chen:92,picco:97,berche:03a}. It turns out to be difficult to
determine the exponents with sufficient precision to arrive at decisive
statements, but the most likely situation appears to be that $\nu \approx 1$
independent of $q$, while the magnetic ratio $\beta/\nu$ changes with $q$, a
scenario that has recently also found additional support in perturbation theory
\cite{delfino:17}.

A fertile testing ground for predictions relating to the behavior of first-order
transitions under the influence of quenched disorder is the Blume-Capel
model~\cite{blume66,capel66}. It has been used to describe the prime nuclear fuel
uranium dioxide~\cite{blume66}, Mott insulators~\cite{kudin2002,lanata2017},
$^{3}$He--$^{4}$He mixtures~\cite{blume1971,lawrie84} and more general
multi-component fluids~\cite{wilding96}, as well as potentially the hardest
piezomagnet known~\cite{jaime17}. The pure system features a tricritical point
separating second-order and first-order lines of transitions~\cite{zierenberg17}.
There are many open questions concerning the behavior of this model in presence of
quenched disorder, be it of random-bond type as considered here or in the form of
random (crystal) fields~\cite{sumedha17,santos18}. In particular, conflicting results
have been found for the universality class of the ex-first-order segment of the
transition line~\cite{theodorakis12}, and some authors have favored a scenario that
contradicts universality~\cite{malakis09,malakis10}.  In the following, we present
the results of high-statistics Monte Carlo simulations that demonstrate that the
transitions in the second-order and the first-order segments of the transition line
of the pure system are in the same universality class after coupling to the disorder,
and this class is that of the two-dimensional (random) Ising model. Hence any doubts
about the universality of critical behavior in this system are dispelled.

\begin{figure}[tb]
\includegraphics{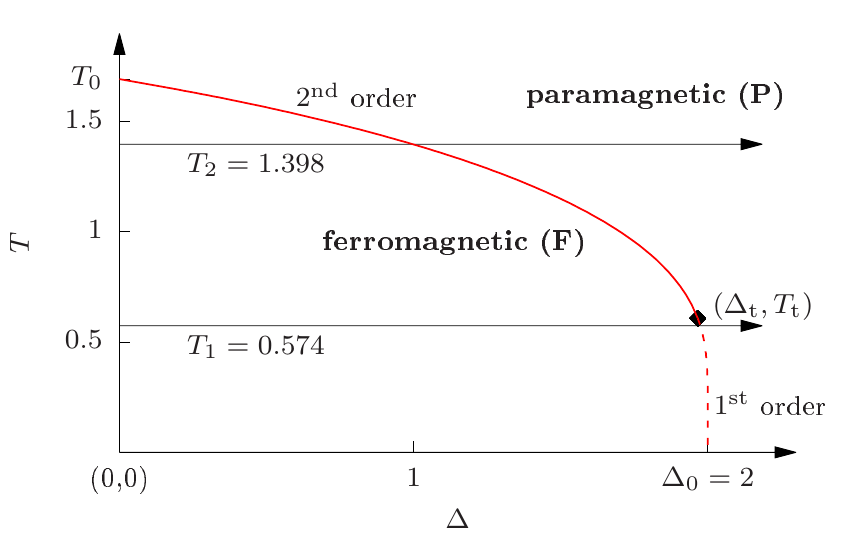}
\caption{(Color online) Phase diagram of the pure two-dimensional
Blume-Capel model~\cite{zierenberg17}, showing the ferromagnetic
(\textbf{F}) and para\-magnetic (\textbf{P}) phases that are
separated by a continuous transition for small $\Delta$ (solid
line) and a first-order transition for large $\Delta$ (dotted
line). The line segments meet at a tricritical point, as indicated
by the black rhombus. The horizontal arrows indicate the paths of
crossing the phase boundary implemented in the simulations of the
present work. \label{fig1}}
\end{figure}

We study the spin-1 or Blume-Capel model~\cite{blume66,capel66} with Hamiltonian
\begin{equation}\label{eq:Ham_dis}
  \mathcal{H}
  = -\sum_{\langle xy \rangle}J_{xy}\sigma_{x}\sigma_{y}+\Delta \sum_{x}\sigma_{x}^{2} \
  = E_J + \Delta E_\Delta = E,
\end{equation}
where the spin variables $\sigma_x\in\{-1,0,+1\}$ live on a square lattice with
periodic boundaries and $\langle xy\rangle$ indicates summation over nearest
neighbors. The couplings $J_{xy}$ are drawn from a bimodal distribution
\begin{equation} \label{eq:bimodal}
  \mathcal{P}(J_{xy})=\frac{1}{2}~[\delta(J_{xy}-J_{1})+\delta(J_{xy}-J_{2})],
\end{equation}
where following Refs.~\cite{malakis09,malakis10} we choose $J_{1}+J_{2} = 2$ and
$J_{1}>J_{2}>0$, so that $r=J_{2} / J_{1}$ defines the disorder strength. The crystal
field $\Delta$ controls the density of vacancies, \emph{i.e.}, sites with
$\sigma_x = 0$. The pure model has been studied extensively (for a review see
Ref.~\cite{zierenberg17}), the phase diagram in the ($\Delta$, $T$)--plane is shown
in Fig.~\ref{fig1}: For small $\Delta$ there is a line of continuous transitions
between the ferromagnetic and paramagnetic phases that crosses the $\Delta = 0$ axis
at $T_{0}\approx 1.693$~\cite{malakis10}. For large $\Delta$, on the other hand, the
transition becomes discontinuous and it meets the $T=0$ line at
$\Delta_0 = zJ/2$~\cite{capel66}, where $z = 4$ is the coordination number (here we
set $J = 1$, and also $k_{\rm B} = 1$, to fix the temperature scale). The two line
segments meet in a tricritical point estimated to be at
$(\Delta_{\rm t}\approx1.966,T_{\rm t}\approx0.608)$~\cite{kwak2015,jung17}. It is
well established that the second-order transitions belong to the universality class
of the two-dimensional Ising model~\cite{zierenberg17}. As $\alpha=0$ there, the
Harris criterion is inconclusive, but explicit studies of the Ising model indicate
that the singularity is only logarithmically
modified~\cite{dotsenko:83,dotsenko81,selke98}. The first-order transition gets
stronger as $\Delta_0$ is approached and, in fact, the interface tension increases
linearly with decreasing temperature~\cite{jung17}. According to the rigorous result
of Aizenman and Wehr~\cite{aizenman:89a}, the transitions must soften under the
presence of even arbitrarily weak disorder, and we expect a second-order transition
to emerge in this regime too.

\begin{figure}[tb]
\includegraphics{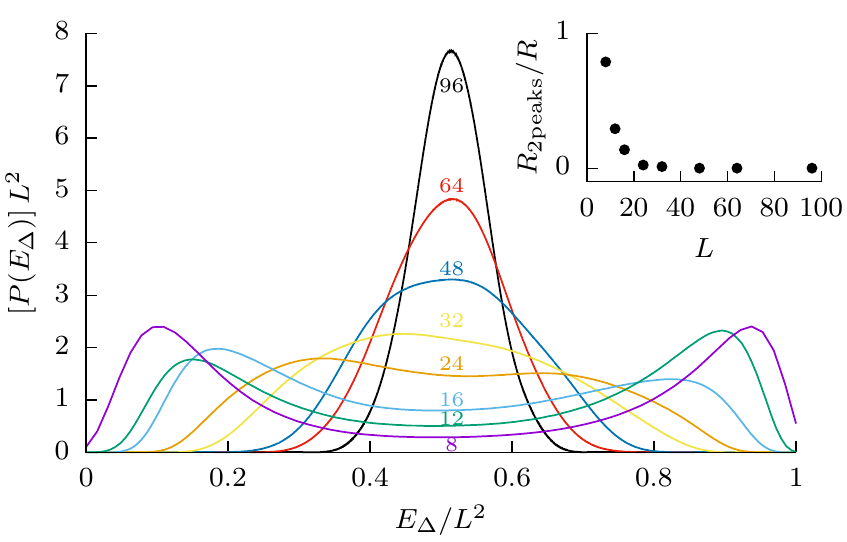}
\caption{(Color online) Probability distribution of crystal fields
$E_\Delta$ of the random-bond Blume-Capel model at $T=0.574$ and
with disorder strength $r=0.6$. The data are averaged over $R=256$
disorder samples. With increasing system size, the double peak
expected for a first-order transition changes to a single broad
peak typical of a continuous transition. The inset shows the
fraction of disorder samples exhibiting a double
peak.\label{fig2}}
\end{figure}

As the phase boundary in the first-order regime is almost vertical, it is most
convenient to cross it by varying the crystal field $\Delta$ while keeping the
temperature constant.  To this end we used a previously developed implementation of
the multicanonical method~\cite{berg91,janke92} applied only to the crystal-field
energy $E_\Delta$ of Eq.~\eqref{eq:Ham_dis} \cite{zierenberg15}. The method
iteratively yields a flat histogram along $E_\Delta$ by replacing the canonical Monte
Carlo weights $\exp(-\beta E)$ by $\exp(-\beta E_J)W(E_\Delta)$ and adapting
$W(E_\Delta)$. Our calculations are implemented in a parallel fashion following the
scheme discussed in Refs.~\cite{zierenberg13,zierenberg15,gross:17}. This procedure
allows us to directly study the probability distribution of $E_\Delta$. The
corresponding histogram for $T = T_1 = 0.574$ close to the transition point, averaged
over $R = 256$ realizations of the random couplings for $r=0.6$, is shown in
Fig.~\ref{fig2}. For small system sizes there is a clear double-peak structure,
characteristic of a first-order phase transition. However, with increasing system
size the distribution changes, exhibiting only a single, symmetric peak, clearly
illustrating the second-order nature of the transition in the limit $L\to\infty$. In
fact, the inset shows that the fraction of disorder samples with a double peak
quickly decays to zero for increasing $L$, with $R_{\rm 2 peaks}/R\approx 0$ for
$L \geq L^{\ast} \approx 32$. This is clear evidence that bond disorder with $r=0.6$
changes the pure first-order phase transition for $T=0.574$ into a disorder-induced
continuous one, yet, with a crossover behavior for small system sizes.

To reveal the universality class of the continuous transition resulting from the
softening by disorder, we used an additional array of canonical Monte Carlo
simulations, employing a combination of a Wolff single-cluster
update~\cite{Wolff1989} of the $\pm 1$ spins and a single-spin flip Metropolis
update~\cite{Blote95,hasen2010,malakis12,zierenberg17}. We restricted these
simulations to the two temperature points indicated by the arrows in Fig.~\ref{fig1},
the case $T_1 = 0.574$ crossing the phase boundary in the first-order regime, and the
choice $T_2 = 1.398$ in the second-order part of the transition
line~\cite{malakis10,zierenberg17}. Using this approach, we simulated the system
sizes $L\in \{8, 12,16,24,32,48,64,96,128,192,256\}$ for disorder strength $r=0.6$.
The ensemble sizes, $R$, of disorder realizations used are as follows:
$R = 5\times 10^3$ for $L = 8 - 32$, $R = 3\times 10^3$ for $L = 48 - 96$, and
$R = 1\times 10^3$ for $L > 96$. Error bars were computed from the sample-to-sample
fluctuations.
\begin{figure}[tb]
\includegraphics{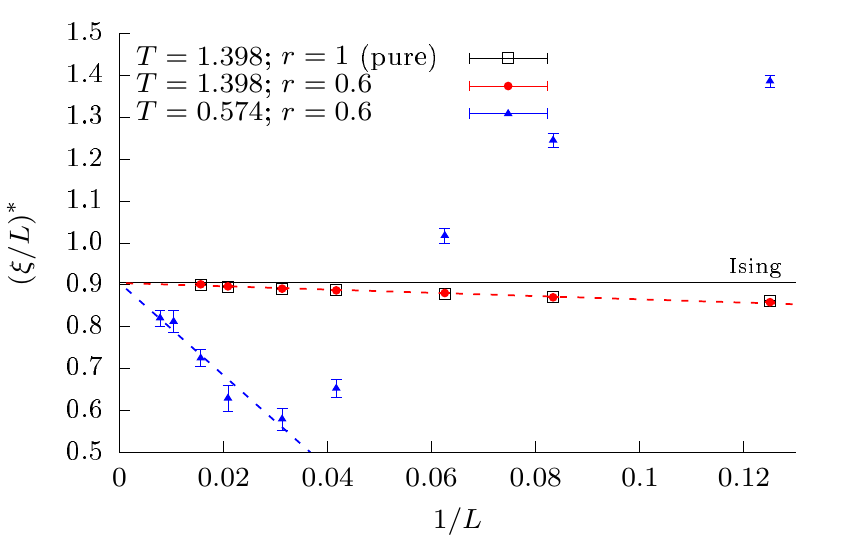}
\caption{(Color online) Finite-size scaling of the
correlation-length ratios at their crossing points,
$(\xi/L)^{\ast}$, for the pure and random model and the two
temperatures considered in this work. Results are shown for the
following pairs ($L$, $2L$) of system sizes: $(8,16)$, $(12,24)$,
$(16,32)$, $(24,48)$, $(32,64)$, $(48,96)$, $(64,128)$,
$(96,192)$, and $(128,256)$. The horizontal line shows the
asymptotic value for the square-lattice Ising model with periodic
boundaries according to Eq.~\eqref{eq:ratio-exact}. The colored
dashed lines show linear fits in $1/L$. \label{fig3}}
\end{figure}

We first discuss the ratio of correlation length and system size, $\xi/L$. This is
known to be universal for a given choice of boundary conditions and aspect ratio. For
Ising spins on a square lattice with periodic boundary conditions as $L \to \infty$
it approaches the value~\cite{salas2000}
\begin{equation} \label{eq:ratio-exact}
  \left(\frac{\xi}{L}\right)_{\infty} = 0.905\,048\,829\,2(4).
\end{equation}
The behavior of the pure, square-lattice Blume-Capel model in the
second-order regime is found to be perfectly consistent with
Eq.~\eqref{eq:ratio-exact}~\cite{zierenberg17}. To determine
$\xi/L$, we use the second-moment definition of the correlation
length $\xi$~\cite{cooper1989, ballesteros2001}. From the Fourier
transform of the spin field, $\hat{\sigma}(\mathbf{k}) = \sum_{\bf
x}\sigma_{\bf x}\exp(i{\bf kx})$, we determine $F = \left\langle
  |\hat{\sigma}(2\pi/L,0)|^2+|\hat{\sigma}(0,2\pi/L)|^2\right\rangle/2$
and obtain the correlation length via~\cite{ballesteros2001}
\begin{equation}\label{eq:xi}
 \xi  \equiv  \frac{1}{2\sin(\pi/L)}\sqrt{\frac{\langle M^2\rangle}{F}-1},
\end{equation}
where $M = \sum_{x}\sigma_{x}$. To estimate the limiting value of $\xi / L$ we relied
on the quotients method~\cite{night,bal96,fytasRFIM}: The crystal-field value where
$\xi_{2L} / \xi_{L} = 2$, \emph{i.e.}, where the curves of $\xi / L$ for the sizes
$L$ and $2L$ cross, defines the finite-size pseudo-critical points
$\Delta^{\rm cross}$. Let us denote the value of $\xi / L$ at these crossing points
as $(\xi/L)^{\ast}$.  In Fig.~\ref{fig3} we show results of $(\xi/L)^{\ast}$ for
three cases, namely the pure and random model at $T = 1.398$ and the random model at
$T = 0.574$. The data for the pure case have been taken from
Ref.~\cite{zierenberg17}, and the horizontal line shows the asymptotic value for the
Ising model with periodic boundaries, cf.\ Eq.~\eqref{eq:ratio-exact}.

In the second-order regime of the pure model, for $T = 1.398$, the
effect of the random bonds is extremely weak for $r=0.6$, with the
results for $(\xi/L)^{\ast}$ practically falling on top of the
data for the pure system. For stronger disorder $r \rightarrow 0$
we expect numerically more pronounced effects, but no
qualitatively different behavior. As is apparent from the data in
Fig.~\ref{fig3}, the results for the disordered and pure systems
have consistent limiting values for $L \to \infty$. For the pure
Blume-Capel model at the same temperature, it was previously found
that $(\xi / L)_{\infty} = 0.906(2)$~\cite{zierenberg17},
perfectly compatible with Eq.~\eqref{eq:ratio-exact}. For the
disordered case a linear fit in $1/L$ for $L \geq 12$ (as shown by
the red dashed line) yields
\begin{equation}\label{eq:ratio_high_T}
  \left(\frac{\xi}{L}\right)_{\infty, \rm random}^{T = 1.398} =
  0.905(2),
\end{equation}
with goodness-of-fit parameter $Q \approx 0.3$. This is clearly consistent with the
Ising value~\eqref{eq:ratio-exact}. An additional analysis of the scaling behavior of
the magnetic susceptibility and specific heat (not shown) is also consistent with
Ising universality, in line with previous analyses~\cite{malakis09,malakis10}.

We now turn to the temperature point $T_1 = 0.574$ in the first-order regime of the
pure model. As it can be seen from the data of Fig.~\ref{fig3} the effect of disorder
is very strong there, leading to huge and non-monotonous scaling corrections. For
smaller lattice sizes, the ratios $(\xi/L)^\ast$ do not show any tendency of
converging to the universal Ising value until $L\approx 32$, when $(\xi/L)^\ast$
attains a minimum. Only for larger lattices the correlation length ratios start to
approach the Ising limit approximately linearly in $1/L$. Taking lattice sizes
$L \geq L^{\ast} \approx 32$ into account, a linear fit in $1 / L$ (as shown by the
blue dashed line) yields
\begin{equation}\label{eq:ratio_low_T}
  \left(\frac{\xi}{L}\right)_{\infty, \rm random}^{T = 0.574} = 0.905(22),
\end{equation}
with $Q \approx 0.3$. The limit is again fully consistent with the Ising value. Note
that the point $L \approx 32$ of the minimum corresponds to the crossover length
scale $L^{\ast}$ determined already as the size where the first-order nature of the
transition disappears for the chosen disorder strength $r = 0.6$, see
Fig.~\ref{fig2}.

\begin{figure}
\begin{tikzpicture}
\node (a) at (0cm, 6cm) {\includegraphics{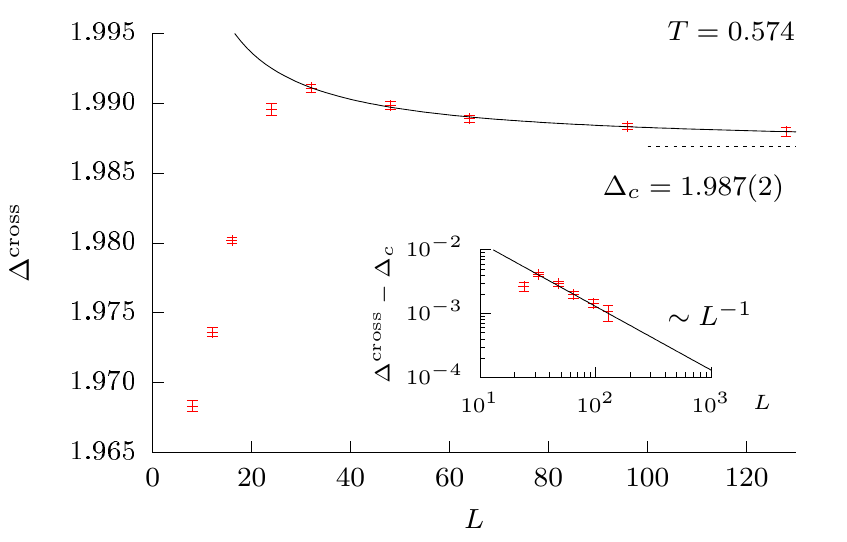}}; \node (b)
at (0cm, 0cm) {\includegraphics{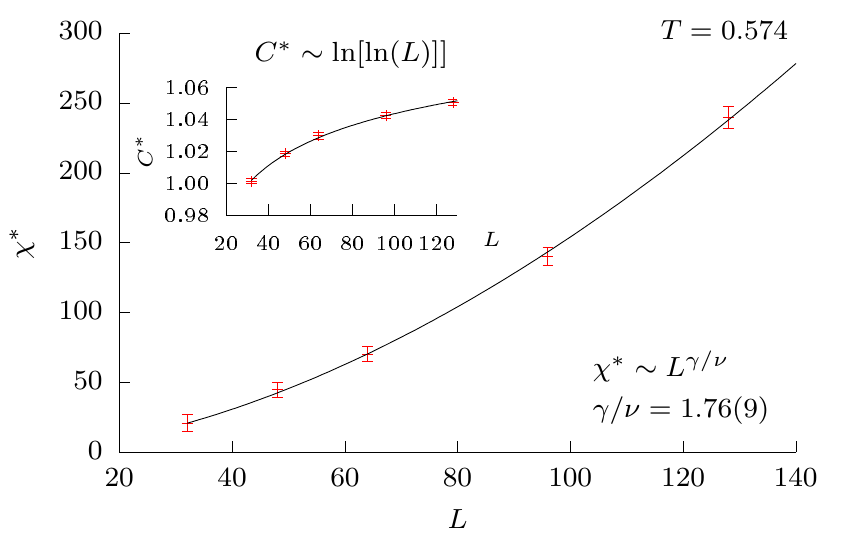}}; \node [above
left=-0.8cm of a] {(a)}; \node [above left=-0.8cm of b] {(b)};
\end{tikzpicture}
\caption{(Color online) Finite-size scaling in the ex-first-order
regime of the Blume-Capel model. (a): Shift behavior of the
pseudo-critical points $\Delta^{\rm cross}$ estimated at the ($L$,
$2L$) crossings of the ratio $\xi / L$ shown in Fig.~\ref{fig3}.
(b): Scaling of the magnetic susceptibility
$\chi^{\ast}=\chi(\Delta^{\rm cross})$ (main panel) and specific
heat  $C^{\ast}=C(\Delta^{\rm cross})$ (inset) evaluated at the
pseudo-critical points for the smaller size of the pairs ($L$,
$2L$) considered.\label{fig4}}
\end{figure}

While the extrapolated value~\eqref{eq:ratio_low_T} of the
correlation-length ratio $(\xi/L)^\ast$ is strong evidence for
Ising behavior, universality classes are characterized by the
entirety of their critical exponents and universal amplitude
ratios. We therefore also considered the scaling of the
pseudo-critical points $\Delta^{\rm cross}$, as well as the
magnetic susceptibility $\chi$ and the specific heat
$C$~\cite{definitions}, both evaluated at $\Delta^{\rm cross}$. We
first considered the scaling for the temperature $T_1=0.574$ in
the first-order regime of the pure system. For large system sizes,
the crossing points are expected to scale as
\begin{equation}\label{eq:cross}
  \Delta^{\rm cross} = \Delta_{\rm c} + bL^{-1/\nu},
\end{equation}
where $\nu=1$ for the two-dimensional Ising universality class.
Our data for the pseudo-critical points are shown in
Fig.~\ref{fig4}(a), and we again observe strong scaling
corrections with a pronounced turnaround in the behavior around
the crossover length scale $L^{\ast} \approx 32$. As the inset
illustrates, however, the behavior for $L \ge 32$ is in perfect
agreement with the inversely linear behavior expected from
Eq.~\eqref{eq:cross} with $\nu=1$. In fact, a fit of the
form~\eqref{eq:cross} for $L \geq 32$ with $Q \approx 0.8$ yields
the critical crystal-field value $\Delta_{\rm c} = 1.987(2)$ and
the estimate $\nu = 1.01(27)$.

Our results for the magnetic susceptibility $\chi$ and the specific heat $C$
evaluated at the pseudo-critical points $\Delta^{\rm cross}$ at $T_{1}=0.574$ are
shown in Fig.~\ref{fig4}(b). Following the above analysis, we exclude small system
sizes $L \leq 24$. For the magnetic susceptibility, a fit of the form
$\chi^{\ast} \sim L^{\gamma/\nu}$ yields $\gamma/\nu=1.76(9)$ with $Q \approx 0.9$,
fully compatible within error bars to the Ising value $1.75$. The specific heat,
shown in the inset of Fig.~\ref{fig4}(b), is well described by a double logarithm
$C^{\ast}\sim \ln{[\ln{(L)}]}$ as predicted by Ref.~\cite{dotsenko81}, the
corresponding fit quality being $Q\approx0.9$. Similarly strong corrections to
scaling in susceptibility data have also been reported for the diluted Ising
model~\cite{wang:90}. An analogous analysis of our data at the higher temperature
$T_2=1.398$ in the second-order regime also yields values compatible to the Ising
behavior, but without the strong scaling corrections observed for $T_1=0.574$.

To summarize, we have used the two-dimensional Blume-Capel model to investigate the
effect of quenched bond disorder on originally second- and first-order phase
transitions. We particularly focused on the effects in the originally first-order
regime, a topic that has been controversial in the literature of disordered systems.
We find that the disorder-induced continuous transitions in both segments of the
phase diagram of the model belong to the universality class of the pure Ising
ferromagnet with logarithmic corrections.  This appears to be the physically most
plausible scenario given that both transitions are between the same ferromagnetic and
paramagnetic phases (Fig.~\ref{fig1}), supporting the strong universality
hypothesis~\cite{reis1996, heuer1991,talapov1994}. While the leading behavior of the
disordered system is hence consistent across the full transition line, there are
dramatic differences in the scaling corrections which appear to be minimal for the
originally second-order transition but maximal and non-monotonic for the case of the
originally first-order transition.

Although universality is a cornerstone in the theory of critical phenomena, it stands
on a less solid foundation for the case of systems subject to quenched disorder. An
explicit confirmation of the behavior of disordered models in this respect is
therefore of fundamental importance for the theory as a whole (see also
Ref.~\cite{fytasRFIM}). In this sense the unambiguous findings presented here set the
stage for studies of similar systems in three dimensions, where one expects disorder
to be relevant only beyond a finite
threshold~\cite{hui:89a,berker93,chatelain:01a,chatelain:05}. A better understanding
of the bond-disordered Blume-Capel model in three dimensions should be of relevance
for a range of experimental systems including $^{3}$He--$^{4}$He mixtures in porous
media as well as impurities in uranium dioxide. Finally, when replacing the random
bonds by random fields the Blume-Capel model might hold an answer to the intriguing
question of whether first-order transitions can survive randomness if it couples to
the order parameter instead of to the energy density \cite{cardy:97a,sumedha17}.

\begin{acknowledgments}

N.G.F is grateful to V. Mart\'{i}n-Mayor for his motivating
comments that triggered this work. The project was financially
supported by the Deutsch-Franz\"osische Hochschule (DFH-UFA)
through the Doctoral College ``${\mathbb L}^4$'' under Grant No.\
CDFA-02-07 as well as by the EU FP7 IRSES network DIONICOS under
contract No.\ PIRSES-GA-2013-612707. J.Z. received financial
support from the German Ministry of Education and Research (BMBF)
via the Bernstein Center for Computational Neuroscience (BCCN)
G{\"o}ttingen under Grant No.~01GQ1005B. This research has been
supported by the National Science Centre, Poland, under grant
No.~2015/19/P/ST3/03541. This project has received funding from
the European Union's Horizon 2020 research and innovation
programme under the Marie Sk{\l}odowska--Curie grant agreement No.
665778. This research was supported in part by PLGrid
Infrastructure.

\end{acknowledgments}

{}


\begin{thebibliography}{199}

\bibitem{young:book} A.P. Young, ed., \emph{Spin Glasses and Random Fields} (World
Scientific, Singapore, 1997).

\bibitem{tabei2006} S. M. A. Tabei, M. J. P. Gingras, Y.-J. Kao, P. Stasiak, and J.-Y. Fortin, Phys. Rev. Lett. \textbf{97}, 237203 (2006).
\bibitem{silevitch2007} D. M. Silevitch, D. Bitko, J. Brooke, S. Ghosh, G. Aeppli, and T. F. Rosenbaum, Nature \textbf{448}, 567 (2007).
\bibitem{maritan1994} A. Maritan, M. Cieplak, T. Bellini, and J. R. Banavar, Phys. Rev. Lett.  \textbf{72}, 4113 (1994).
\bibitem{carlson2006} E. W. Carlson, K. A. Dahmen, E. Fradkin, and S. A.  Kivelson, Phys. Rev. Lett.  \textbf{96}, 097003 (2006)
\bibitem{nagaosa2010} N. Nagaosa, J. Sinova, S. Onoda, A. H. MacDonald, and N.  P. Ong, Rev. Mod. Phys. \textbf{82}, 1539 (2010).
\bibitem{vojta2014} T. Vojta and J. A. Hoyos, Phys. Rev. Lett. \textbf{112}, 075702 (2014).
\bibitem{smith2016} J. Smith, A. Lee, P. Richerme, B. Neyenhuis, P. W. Hess, P. Hauke, M. Heyl, D. A. Huse, and C. Monroe, Nat. Phys. \textbf{12}, 907 (2016).
\bibitem{barghathi2012} H. Barghathi and T. Vojta, Phys. Rev. Lett. {\bf 109}, 170603 (2012).
\bibitem{harris74} A.B. Harris, J. Phys. C {\bf 7}, 1671 (1974).

\bibitem{weinrib:83} A. Weinrib and B.I. Halperin, Phys. Rev. B {\bf 27}, 413 (1983).

\bibitem{luck:93a} J.M. Luck, Europhys. Lett. {\bf 24}, 359 (1993).

\bibitem{wj:04a} W. Janke and M. Weigel, Phys. Rev. B {\bf 69}, 144208 (2004).

\bibitem{barghathi:14} H. Barghathi and T. Vojta, Phys. Rev. Lett. {\bf 113}, 120602 (2014).

\bibitem{fricke:17} N. Fricke, J. Zierenberg, M. Marenz, F.P. Spitzner, V.
Blavatska, and W. Janke, Condens. Matter Phys. {\bf 20}, 13004
(2017).

\bibitem{fricke:17b} J. Zierenberg, N. Fricke, M. Marenz, F.P. Spitzner, V.
Blavatska, and W. Janke, Phys. Rev. E {96}, 062125 (2017).

\bibitem{dotsenko:83} V.S. Dotsenko and V.S. Dotsenko, Adv. Phys. {\bf 32}, 129 (1983).

\bibitem{shalaev:84} B.N. Shalaev, Sov. Phys. Solid State {\bf 26}, 1811 (1984); Phys. Rep. {\bf 237}, 129 (1994).

\bibitem{shankar:87} R. Shankar, Phys. Rev. Lett. {\bf 58}, 2466 (1987); \emph{ibid.} {\bf 61}, 2390 (1988).

\bibitem{ludwig:88} A.W.W. Ludwig, Phys. Rev. Lett. {\bf 61}, 2388 (1988); Nucl. Phys B {\bf 330}, 639 (1990).

\bibitem{wang:90} J.-S. Wang, W. Selke, V.S. Dotsenko, and V.B. Andreichenko, Physica A {\bf 164}, 221 (1990).

\bibitem{hasenbusch:08a} M. Hasenbusch, F.P. Toldin, A. Pelissetto, and E. Vicari, Phys. Rev. E {\bf 78}, 011110 (2008).

\bibitem{kenna:08} R. Kenna and J.J. Ruiz-Lorenzo, Phys. Rev. E {\bf 78}, 031134 (2008).

\bibitem{dotsenko:17} V. Dotsenko, Y. Holovatch, M. Dudka, and M. Weigel, Phys. Rev. E {\bf 95}, 032118 (2017).

\bibitem{cardy:99a} J.L. Cardy, Physica A {\bf 263}, 215 (1999).

\bibitem{aizenman:89a} M. Aizenman and J. Wehr, Phys. Rev. Lett. {\bf 62}, 2503 (1989).

\bibitem{hui:89a} K. Hui and A.N. Berker, Phys. Rev. Lett. {\bf 62}, 2507 (1989).

\bibitem{berker93} A.N. Berker, Physica A {\bf 194}, 72 (1993).

\bibitem{cardy:97a} J.L. Cardy and J.L. Jacobsen, Phys. Rev. Lett. {\bf 79}, 4063 (1997).

\bibitem{goswami:08} P. Goswami, D. Schwab, and S. Chakravarty, Phys. Rev. Lett. {\bf 100}, 015703 (2008).

\bibitem{greenblatt:09} R.L. Greenblatt, M. Aizenman, and J.L. Lebowitz, Phys.
  Rev. Lett. {\bf 103}, 197201 (2009); Physica A {\bf 389}, 2902 (2010).

\bibitem{hrahsheh:12} F. Hrahsheh, J.A. Hoyos, and T. Vojta, Phys. Rev. B {\bf 86}, 214204 (2012).

\bibitem{vojta:06} T. Vojta, J. Phys. A: Math. Gen. {\bf 39}, R143 (2006).

\bibitem{martin:14} P. Villa Mart\'{i}n, J.A. Bonachela, and M.A. Mu\~{n}oz, Phys. Rev. E {\bf 89}, 012145 (2014).

\bibitem{bellafard:12} A. Bellafard, H.G. Katzgraber, M. Troyer, and S. Chakravarty, Phys. Rev. Lett. {\bf 109}, 155701 (2012).

\bibitem{zhu:15} Q. Zhu, X. Wan, R. Narayanan, J.A. Hoyos, and T. Vojta, Phys. Rev. B {\bf 91}, 224201 (2015).

\bibitem{chatelain:01a} C. Chatelain, B. Berche, W. Janke, and P.E. Berche, Phys. Rev. E {\bf 64}, 036120 (2001).

\bibitem{chatelain:05} C. Chatelain, B. Berche, W. Janke, and P.-E. Berche, Nucl. Phys. B {\bf 719}, 275 (2005).

\bibitem{chen:92} S. Chen, A.M. Ferrenberg, and D.P. Landau, Phys. Rev. Lett. {\bf 69}, 1213 (1992).

\bibitem{picco:97} M. Picco, Phys. Rev. Lett. {\bf 79}, 2998 (1997).

\bibitem{berche:03a} B. Berche and C. Chatelain, in \emph{Order, Disorder And
Criticality}, edited by Y. Holovatch (World Scientific, Singapore,
2004), p. 147.

\bibitem{delfino:17} G. Delfino, Phys. Rev. Lett. {\bf 118}, 250601 (2017); G. Delfino and E. Tartaglia, J. Stat. Mech. (2017) 123303.

\bibitem{blume66} M. Blume, Phys. Rev. {\bf 141}, 517 (1966).

\bibitem{capel66} H.W. Capel, Physica (Utr.) {\bf 32}, 966 (1966); {\bf 33}, 295 (1967); {\bf 37}, 423 (1967).

\bibitem{kudin2002} K.~N. Kudin, G. E. Scuseria, and R.~L. Martin, Phys. Rev. Lett. {\bf 89}, 266402 (2002).
  
\bibitem{lanata2017} N. Lanat{\`a}, Y. Yao, X. Deng, V. Dobrosavljevi{\'c}, and G. Kotliar, Phys. Rev. Lett. {\bf 118}, 126401 (2017).
  
\bibitem{jaime17} M. Jaime, \emph{et al.}, Nat. Commun. {\bf 8}, 99 (2017).

\bibitem{blume1971} M. Blume, V.~J. Emery, and R.~B. Griffiths, Phys. Rev. A {\bf 4}, 1071 (1971).

\bibitem{lawrie84} I.D. Lawrie and S. Sarbach, in \emph{Phase Transitions and Critical
Phenomena}, edited by C. Domb, J.L. Lebowitz., Vol. 9 (Academic
Press, London, 1984).

\bibitem{wilding96} N.B. Wilding, Phys. Rev. E {\bf 53}, 926 (1996).

\bibitem{zierenberg17}  J. Zierenberg, N.G. Fytas, M. Weigel, W. Janke, and A. Malakis, Eur. Phys. J. Special Topics {\bf 226}, 789 (2017).

\bibitem{sumedha17} Sumedha and N.K. Jana, J. Phys. A: Math. Theor. {\bf 50}, 015003 (2017).

\bibitem{santos18} P.V. Santos, F.A. da Costa, and J.M. de Ara\'{u}jo, J. Magn. Magn. Mater. {\bf 451}, 737 (2018).

\bibitem{theodorakis12} P.E. Theodorakis and N.G. Fytas, Phys. Rev. E {\bf 86}, 011140 (2012).

\bibitem{malakis09} A. Malakis, A.N. Berker, I.A. Hadjiagapiou, and N.G. Fytas, Phys. Rev. E {\bf 79}, 011125 (2009).

\bibitem{malakis10} A. Malakis, A.N. Berker, I.A. Hadjiagapiou, N.G. Fytas, and T. Papakonstantinou, Phys. Rev. E {\bf 81}, 041113 (2010).

\bibitem{kwak2015} W. Kwak, J. Jeong, J. Lee, and D.-H. Kim, Phys. Rev. E {\bf 92}, 022134 (2015).

\bibitem{jung17}  M. Jung and D.-H. Kim, Eur. Phys. J. B {\bf 90}, 245 (2017).

\bibitem{dotsenko81} V.S. Dotsenko and V.S. Dotsenko, Sov. Phys. JETP Lett. {\bf 33}, 37 (1981).

\bibitem{selke98} W. Selke, L.N. Shchur, and O.A. Vasilyev, Physica A {\bf 259}, 388 (1998).

\bibitem{berg91} B.A. Berg and T. Neuhaus, Phys. Lett. B {\bf 267}, 249 (1991); Phys. Rev. Lett. {\bf 68}, 9 (1992).

\bibitem{janke92} W. Janke, Int. J. Mod. Phys. C {\bf 03}, 1137 (1992); Physica A {\bf 254}, 164 (1998).

\bibitem{zierenberg15} J. Zierenberg, N.G. Fytas, and W. Janke, Phys. Rev. E {\bf 91}, 032126 (2015).

\bibitem{zierenberg13} J. Zierenberg, M. Marenz, and W. Janke, Comput. Phys. Commun. {\bf 184}, 1155 (2013).

\bibitem{gross:17} J. Gross, J. Zierenberg, M. Weigel, and W. Janke, Comput. Phys. Commun. {\bf 224}, 387 (2018).

\bibitem{Wolff1989} U. Wolff, Phys. Rev. Lett. {\bf 62}, 361 (1989).

\bibitem{Blote95} H.W.J. Bl\"{o}te, E. Luijten, and J.R. Heringa, J. Phys. A: Math. Gen. {\bf 28}, 6289 (1995).

\bibitem{hasen2010} M. Hasenbusch Phys. Rev. B {\bf 82}, 174433 (2010).

\bibitem{malakis12} A. Malakis, A.N. Berker, N.G. Fytas, and T. Papakonstantinou, Phys. Rev. E {\bf 85}, 061106 (2012).

\bibitem{salas2000} J. Salas, and A.D. Sokal, J. Stat. Phys. {\bf 98}, 551 (2000).

\bibitem{cooper1989} F. Cooper, B. Freedman, and D. Preston, Nucl. Phys. B {\bf 210}, 210 (1982).

\bibitem{ballesteros2001} H.G. Ballesteros, L.A. Fern{\'a}ndez, V. Mart{\'i}n-Mayor, A.
Mu{\~n}oz Sudupe, G. Parisi, and J.J. Ruiz-Lorenzo, J. Phys. A:
Math. Gen. {\bf 32}, 1 (1999).

\bibitem{night} M.P. Nightingale, Physica (Amsterdam) {\bf 83A}, 561 (1976).

\bibitem{bal96} H.G. Ballesteros, L.A. Fern\'{a}ndez, V. Mart\'{i}n-Mayor, and A. Mu\~{n}oz-Sudupe, Phys. Lett. B {\bf 378}, 207 (1996).

\bibitem{fytasRFIM}  N.G. Fytas and V. Mart\'{i}n-Mayor, Phys. Rev. Lett. {\bf 110}, 227201
(2013); Phys. Rev. E {\bf 93}, 063308 (2016); N.G. Fytas, V.
Mart\'{i}n-Mayor, M. Picco, and N. Sourlas, Phys. Rev. Lett. {\bf
116}, 227201 (2016); Phys. Rev. E {\bf 95}, 042117 (2017).

\bibitem{definitions} Following Refs.~\cite{zierenberg15,zierenberg17} we have used the following
definitions for the magnetic susceptibility and the specific heat:
$\chi = \beta\left(\langle M^2 \rangle-\langle
|M|\rangle^2\right)/L^2$ and $C \equiv \frac{\partial \langle E_J
\rangle}{\partial\Delta}\frac{1}{L^2} = - \beta \left(\langle E_J
E_{\Delta}\rangle - \langle E_J\rangle\langle
E_{\Delta}\rangle\right)/L^2$, respectively.

\bibitem{reis1996} F.D.A. Aar{\~a}o Reis, S.L. de Queiroz, and R.R. dos Santos, Phys. Rev. B {\bf 54}, R9616 (1996).

\bibitem{heuer1991} H.-O. Heuer, Europhys. Lett. {\bf 16}, 503, (1991).

\bibitem{talapov1994} A.L. Talapov and L.N. Shchur, Europhys. Lett. {\bf 27}, 193 (1994).

\end{thebibliography}
\end{document}